\documentclass{nature}[a4paper]
\usepackage{bm}
\usepackage{pifont}
\usepackage{marvosym}
\usepackage[colorlinks,linkcolor = blue,citecolor = blue,anchorcolor = blue]{hyperref}
\usepackage{multirow}
\usepackage{geometry}
\usepackage{authblk}
\usepackage{amsmath}
\usepackage{graphicx}
\usepackage{float}
\usepackage{xcolor}
\makeatletter
\let\saved@includegraphics\includegraphics
\AtBeginDocument{\let\includegraphics\saved@includegraphics}
\renewenvironment*{figure}{\@float{figure}}{\end@float}
\makeatother
\usepackage[labelfont=bf]{caption}
\usepackage{cite}
\usepackage{soul}

\title{Doubled Shapiro steps in a dynamic axion insulator Josephson junction\footnotetext{$^\ast$  E-mail: jianghuaphy@fudan.edu.cn, xcxie@pku.edu.cn}}

\author[1,2,3]{Yu-Hang Li}
\author[4]{Ziqian Zhou}
\author[3,5]{Ran Cheng}
\author[6\ \Letter]{Hua Jiang}
\author[4,6,7\ \Letter]{X. C. Xie}

\affil[1]{School of Physics, Nankai University, Tianjin 300071, China}
\affil[2]{State Key Laboratory of the Surface Physics and Department of Physics, Fudan University, Shanghai 200433, P. R. China}
\affil[3]{Department of Electrical and Computer Engineering, University of California, Riverside, California 92521, USA}
\affil[4]{International Center for Quantum Materials, School of Physics, Peking University, Beijing 100871, China}
\affil[5]{Department of Physics and Astronomy, University of California, Riverside, California 92521, USA}
\affil[6]{Interdisciplinary Center for Theoretical Physics and Information Sciences (ICTPIS), Fudan University, Shanghai 200433, China}
\affil[7]{Hefei National Laboratory, Hefei 230088, China}

\graphicspath{{figures/}}

\date{}                   
\begin{document}
\sloppy
	
	\maketitle
	
	\vspace{10pt}

\begin{abstract}
\textbf{Abstract: Dynamic axion insulators feature a time-dependent axion field that can be induced by antiferromagnetic resonance. Here, we show that a Josephson junction incorporating this dynamic axion insulator between two superconductors exhibits a striking doubled Shapiro steps wherein all odd steps are completely suppressed in the jointly presence of a DC bias and a static magnetic field. The resistively shunted junction simulation confirms that these doubled Shapiro steps originate from the distinctive axion electrodynamics driven by the antiferromagnetic resonance, which thus not only furnishes a hallmark to identify the dynamic axion insulator but also provides a method to evaluate its mass term. Furthermore, the experimentally feasible differential conductance is also determined. Our work holds significant importance in condensed matter physics and materials science for understanding the dynamic axion insulator, paving the way for its further exploration and applications.}\\
\end{abstract}

\noindent\textbf{Introduction}\\
The advent of dynamic axion insulator (DAI)~\cite{Peccei1977,Weinberg1978,Wilczek1978,Li2010dynamic} has sparked a surge of interest across multidisciplinary fields including particle physics, cosmology, optics, and particularly condensed matter physics~\cite{tokura2019magnetic,Bhattacharyya2021recent,Bernevig2022progress,WANG2021intrinsic,nenno2020axion}. Because of the inherent dynamic axion field, DAI possesses an additional term $\mathcal{L}_{\theta}=\theta(t)\alpha{\bf{E}}\cdot{\bf{B}}/\pi$ in the Lagrangian~\cite{Li2010dynamic,qi2008topological}, where $\theta(t)$ is the time-dependent, massive axion field, $\alpha$ is the fine structure constant, $\bf{E}$ and $\bf{B}$ are conventional magnetoelectric field. This Lagrangian introduces a magnetoelectric contribution to the Maxwell's equations, making it possible to realize the axion polaritons or axion instability in condensed systems~\cite{Li2010dynamic,Ooguri2012instability,Imaeda2019Axion}, to drive anomalous magnetoelectric transport\cite{Sekine2016chiral,Gooth2019}, to detect the dark matters~\cite{Beck2013possible,BECK2015axion,Klaer2017,Borsanyi2016}, and so on~\cite{Sekine2021axion,Xiao2121Nonlinear}.

\noindent Several candidate systems have been predicted to be DAIs~\cite{Wang2011Topological,wang2016dynamical,wang2019dynamic,Zhu2021tunable,Sekine2016chiral,Wang2020dynamic,Taguchi2018electromagnetic}. In particular, recent first principle calculations suggest that Mn${}_2$Bi${}_2$Te${}_5$ and its family, which possess coexisting antiferromagnetic ordering and topology~\cite{Eremeev2022magnetic}, may be materials of this category~\cite{zhang2020large,Cao2021growth}. Since the axion field in these materials originates from the exchange interaction between topological electrons and the antiparallel magnetic moments, or the N\'eel vectors, the axion dynamics can then be readily achieved through the magnetic fluctuation that stimulates a time-dependent N\'eel vector~\cite{Li2010dynamic,zhang2020large}. However, because of the ultrafast dynamics of the axion field, detecting this novel state remains an outstanding challenge, thereby significantly hindering its further exploration and potential applications. Meanwhile, there is also fierce debate whether the magnitude of the axion mass in condensed matter materials is on the order of eV or meV~\cite{Sekine2014axionic,Jan2021axion,Ishiwata2021axion,Ishiwata2022topology}.

\noindent In this work, we demonstrate that this exotic quantum state can be unambiguously verified by the transport signature of a superconductor-DAI-superconductor Josephson junction. Since DAIs possess spontaneous antiferromagnetic ordering and axion field, an antiferromagnetic resonance (AFMR) can occurs in DAI under the driven of a linear-polarized microwave if the microwave frequency $\omega$ matches its intrinsic AFMR frequency~\cite{kittel1951theory,Li2020,Priyanka2020sub}. In this case the N\'eel vector, hence the magnetic exchange gap in the DAI, become time-dependent, giving rise to a dynamic axion field binding to the AFMR. The quantitative expression for this dynamic axion field turns out to be a sinusidal function possessing a strikingly doubled frequency $\omega_1=2\omega$. Subsequently, this dynamic axion field leads to a magnetoelectric current in the presence of a static magnetic field, which becomes a harmonic supercurrent when the DAI is sandwiched between two superconductors and simultaneously, populates a coherent phase difference possessing the same doubled frequency $\omega_1=2\omega$. On the other hand, a DC bias $V_0$ can also induce a harmonic phase difference across the Josephson junction, whose frequency is proportional to the applied bias voltage is $\omega_2=2eV_0$.
Consequently, when the two frequencies $\omega_1$ and $\omega_2$ are commensurate with each other, such a superconductor-DAI-superconductor Josephson junction exhibits a remarkable doubled Shapiro steps where all odd steps are completely suppressed.
Because the magnetoelectric current originates from the axion electrodynamics driven by the AFMR, those Shapiro steps thus not only furnish a fingerprint of DAIs but also provide a method to evaluate the axion mass.

\noindent\textbf{Results}\\
\noindent\textbf{Effective model of the DAI}\\
\noindent The generic low-energy effective Hamiltonian for a DAI defined on the basis of $\psi_{{\bf{k}}}=\begin{pmatrix}|p^+_z,\uparrow\rangle,&|p^+_z,\downarrow\rangle,&|p^-_z,\uparrow\rangle,&|p^-_z,\downarrow\rangle\end{pmatrix}^{\textrm{T}}$ has the form~\cite{Li2010dynamic,zhang2020large}
\begin{align}\label{DAI_Ham}
\mathcal{H}(t)=\sum_{\alpha=1}^5 d_{\alpha}({\bf{k}})\Gamma^{\alpha},
\end{align}
where $d_{1,2,3,4,5}({\bf{k}})=[Ak_x+m_5n_x(t),Ak_y,Ak_z,m_0+Bk^2,m_5n_z(t)]$ and $\Gamma^{1,2,3,4,5}=[\sigma_x\otimes s_x,\sigma_x\otimes s_y,\sigma_y,\sigma_z,\sigma_x\otimes s_z]$. Here, $A$, $B$, $m_0$ and $m_5$ are system parameters. $n_{x(z)}(t)$ denotes the $x$-component ($z$-component) of the  N\'eel vector, which can be time-dependent in the presence of magnetic fluctuations. $\sigma_{x,y,z}$ and $s_{x,y,z}$ are Pauli matrices acting on the orbital and spin spaces, respectively. The lattice wave vectors $k_{x,y,z}$ are defined in the first Brillouin zone with $k=\sqrt{k_x^2+k_y^2+k_z^2}$. The first four terms in Eq.~\ref{DAI_Ham} describe a topological insulator preserving both parity  $\mathcal{P}=\sigma_z$ and time reversal $\mathcal{T}=i\tau_y\mathcal{K}$ ($\mathcal{K}$ is the complex conjugation operator) symmetry, whose axion field is quantized to $\theta=0$ when $B=0$ while $\theta=\pi$ otherwise~\cite{qi2008topological}. Whereas the fifth term describes the exchange interaction between the topological electron and the dynamic, antiparallel magnetic moments, which breaks these two symmetries explicitly and thus introduces an additional dynamic part to the static axion field, giving rise to the DAI. The system parameters are generally taken as $A=1$, $B=-0.5$, $m_0=0.1$, $m_5=0.1$. However, our theory is universal and not limited to these parameters.

\begin{figure}[t]
  \centering
  \includegraphics[width=1.0\textwidth]{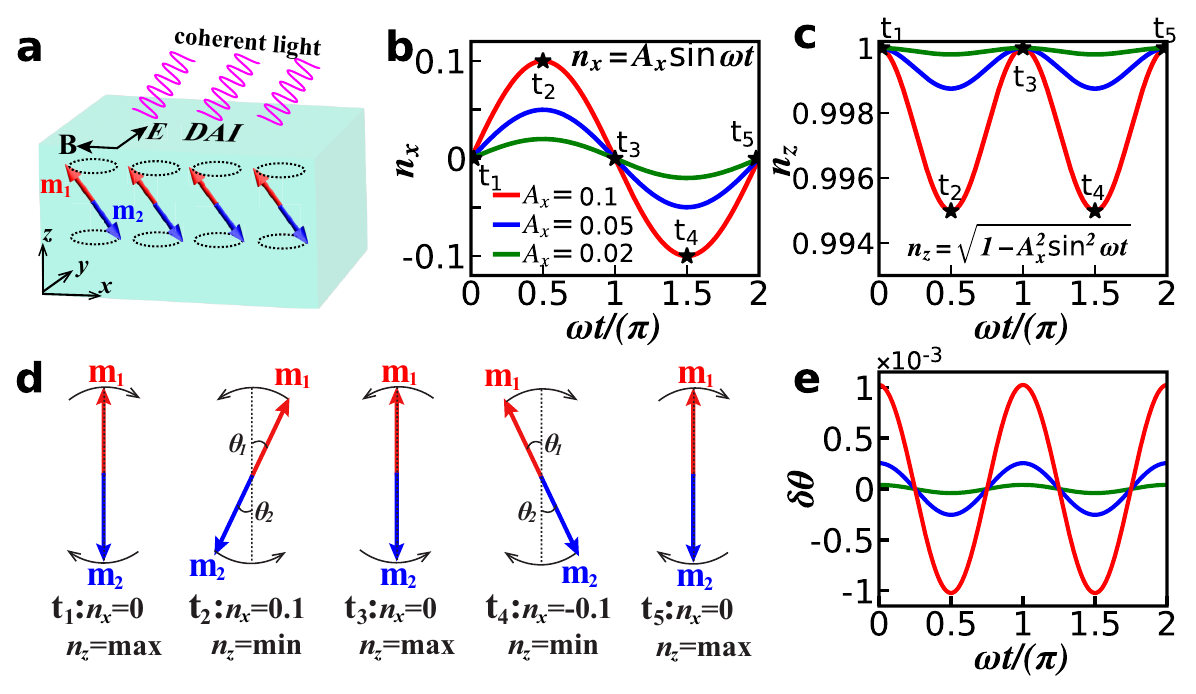}
  \caption{\textbf{Dynamic axion field induced by antiferromagnetic resonance in DAI.}
  		{\textbf{a}} Schematic for a dynamic axion insulator driven by a linearly polarized coherent light. The arrows labeled $\bf{B}$ and $\bf{E}$ represent the magnetic and electric components of the applied light. The red and blue arrows labeled ${\bf{m}}_1$ and ${\bf{m}}_2$ represent the antiparallel magnetic moments forming a N\'eel vector, which oscillates linearly under the driven of the $\bf{B}$ component of the light.
		\textbf{b} and \textbf{c} are temporal evolutions of $n_x$ and $n_z$ under linearly polarized AFMR for three different amplitudes $A_x=0.1$, $0.05$, $0.02$. The black stars marked five representative times ranging from $t_1$ to $t_5$.
		\textbf{d} Schematics for five representative magnetic configurations during a full circle one by one corresponding to the five times marked by black stars in \textbf{b} and \textbf{c}. The colorful arrows represent the magnetic moments ${\bf{m}}_1$ (${\bf{m}}_2$) and $\theta_1$ ($\theta_2$) denotes its polar angle. The black arrow denote the direction of motion of the magnetic moment. The tilting angles of the  N\'eel vector are exaggerated for clarity.
		\textbf{e} Temporal dependence of the dynamic axion field $\delta\theta(t)$ obtained from Eq.~\ref{axion_field} in the adiabatic limit at corresponding AFMR amplitudes. 
		  	}
\label{fig1}
\end{figure}

\noindent\textbf{Antiferromagnetic resonance in the DAI}\\
\noindent Under a linearly polarized microwave, an antiferromagnetic insulator resonates coherently if the frequency of the microwave is identical to its intrinsic AFMR frequency $\omega=2\pi\gamma\sqrt{H_A(2H_E+H_A)}$ given in Supplementary Note 1~\cite{kittel1951theory}, where $\gamma=28$GHz$\cdot$T$^{-1}$ is the gyromagnetic ratio, $H_A$ and $H_E$  are the uniaxial anisotropy and the exchange field between two antiparallel magnetic moments ${\bf{m}}_1$ and ${\bf{m}}_2$. Although the magnetic moment in this case individually rotates clockwise (or counterclockwise) on an elliptical orbit, the N\'eel vector ${\bf{n}}={{\bf{m}}_1}-{{\bf{m}}_2}$ oscillates in a purely linear mode as schematically illustrated in Figs.~\ref{fig1}a and c, traveling back and forth inside a perpendicular plane normal to $\hat{y}$-axis. Consequently, x component of the N\'eel vector has an explicit form $n_x(t)=A_x\sin{\omega t}$ with $A_x$ the AFMR amplitude while z component can therefore be analytically expressed as $n_z(t)=\sqrt{1-A_x^2\sin^2{\omega t}}$ since the magnitude of the N\'eel vector is normalized. Figures~\ref{fig1}b and c show the temporal evoluations of $n_x$ (b) and $n_z$ (c), respectively, at $A_x=0.02$, $0.05$, $0.1$. We see that the periodicity of $n_z$ is halved compared to that of $n_x$, indicating that the frequency of $n_z$ is multipled.

\noindent\textbf{Quantitative expression for the dynamic axion field in the DAI}\\
\noindent To establish a quantitative expression for this dynamic axion field driven by the linearly polarized AFMR, we employ the gauge invariant expression for the axion field in terms of the Chern-Simons 3-form~\cite{qi2008topological}. While the timescale of the AFMR hence $\theta(t)$ is about 7 orders of magnitude larger than the typical electron response time~\cite{Hassan2016}, the electrons inside the DAI can adjust adiabatically to the instantaneous configuration of N\'eel vector, viewing the AFMR as a static magnetic vector almost frozen in time. As a result, we can treat the DAI adiabatically~\cite{Li2021Spin}. As detailed in Supplementary Note 2, this axion field in the adiabatic limit can be calculated by using~\cite{Li2010dynamic} 
\begin{align}\label{axion_field}
\theta=-\frac{1}{4\pi}\int{d{\bf{k}}^3}\frac{2d+d_4}{(d+d_4)^2d^3}d_5\partial_{k_x}{d_1}\partial_{k_y}{d_2}\partial_{k_z}{d_3},
\end{align}
where $d=\sqrt{\sum_{i}d_i^2}$ with $d_i$ presented in Eq.~\ref{DAI_Ham}. Repeating this calculation at different instant of time gives the time-dependent axion field including a static part and a dynamic one, or $\theta(t)=\theta_0+\delta\theta(t)$. Figure~\ref{fig1}f plots $\delta\theta(t)$ as a function of time at different AFMR amplitudes corresponding to Figs.~\ref{fig1}b and c. It turns out that the temporal evolution of $\delta\theta(t)$ is a harmonic function with a doubled frequency compared to the AFMR, that is $\delta\theta(t)=A_\theta\cos{2\omega t}$ where $A_\theta$ is the amplitude proportional to $A_x^2$. These mathematical relations, especially the frequency doubling of the dynamic axion field $\delta{\theta}(t)$, are further confirmed analytically in Supplementary Note 3.

\begin{figure}[t]
  \centering
  \includegraphics[width=1.0\textwidth]{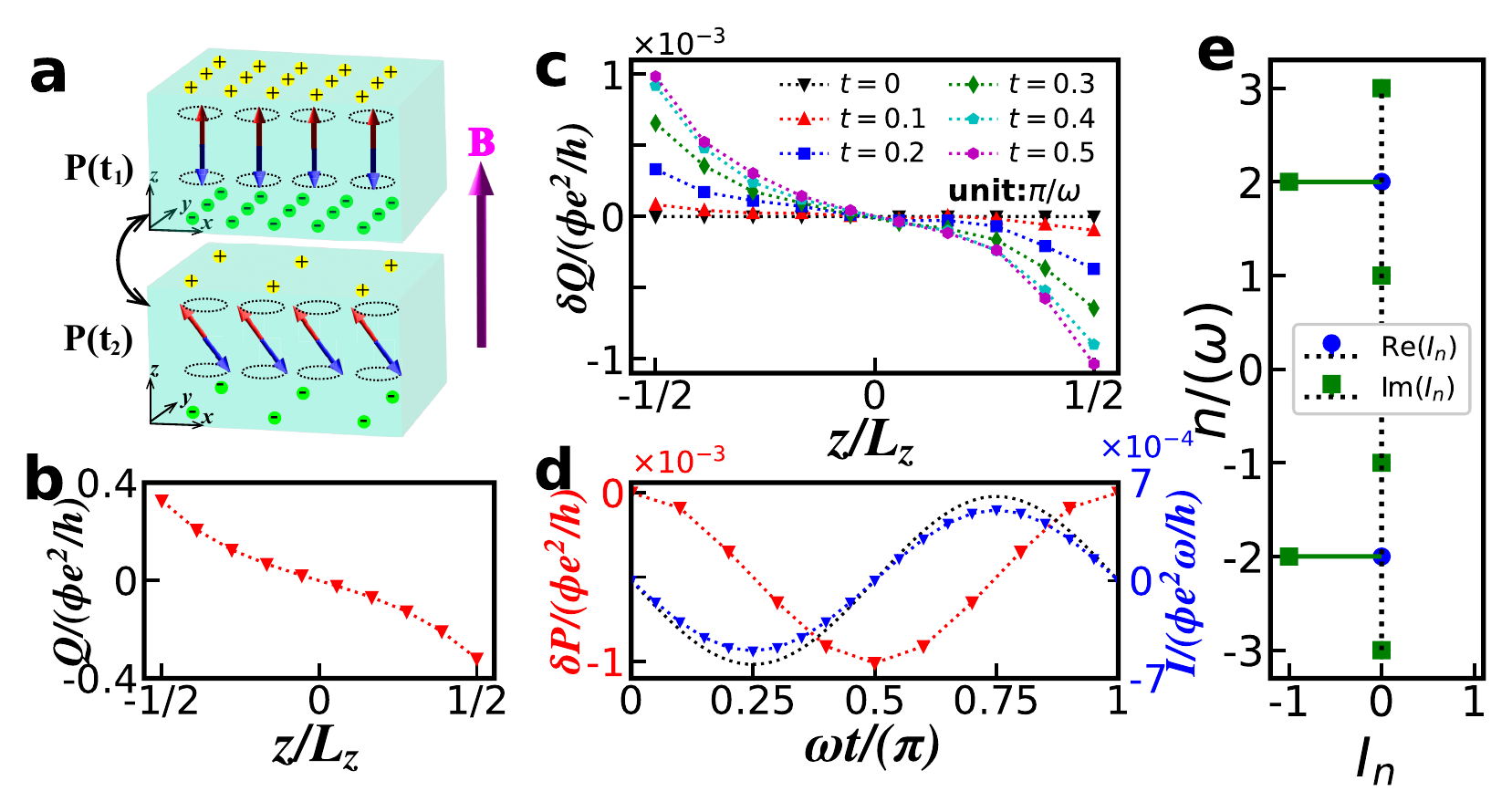}
  \caption{\textbf{Doubled magnetoelectric current in DAI.}
  	\textbf{a} Schematic for the instant charge polarization induced by a static magnetic field $\bf{B}$ along $\hat{z}$-direction at $t_1=0$ and $t_2=2\pi/5\omega$, respectively. The charge polarization driven by AFMR oscillates between these two values in a doubled harmonic mode. The yellow and lime spheres represent positive and negative charge respectively.
	\textbf{b} Unbalanced charge distribution $Q(z)$ at $t_1=0$ for a system with size $L_z=10$ and $L_y=16$. We use periodic boundary condition along $\hat{x}$-direction. Here, $A_x=0.1$ and the magnetic flux inside one unit cell is $\Phi_0=Ba_0^2=0.005\phi_0$ with $\phi_0=h/e$ the magnetic flux quanta.
	\textbf{c} Increment of the charge distribution $\delta Q(z)=Q(z,t)-Q(z,t=0)$ at different times ranging from $t=0$ to $t=0.5\pi.\omega$ with an interval of $0.1\pi/\omega$ at the same parameters as the system in \textbf{b}.
	\textbf{d} Temporal evolutions of the increment of the charge polarization $\delta P=[Q(1/2)-Q(-1/2)]/2$ obtained in the adiabatic limit from the equilibrium Green’s functions provided in Methods (red inverted triangle) and the ensuing dynamic magnetoelectric current calculated from the time derivative of $\delta P$ (blue triangle). The black dash line shows the anticipated magnetoelectric current obtained from the time derivative of instant charge polarization given by Eq.~\ref{axion_field}.
	\textbf{e} Stem plot of the Fourier coefficients $I_n$.
		  	}
\label{fig2}
\end{figure}

\noindent\textbf{Dynamic magnetoelectric current}\\
\noindent  Owing to the magnetoelectric term in the Lagrangian $\mathcal{L}_{\theta}=\theta(t)\alpha{\bf{E}}\cdot{\bf{B}}/\pi$, the Maxwell equation acquires an additional term and thus becomes $\nabla\cdot{\bf{E}}=\rho/\epsilon-\alpha\nabla\theta\cdot{\bf{B}}/\pi$ where $\epsilon$ is the permittivity and $\rho$ is the charge density~\cite{Sekine2021axion}. So a static magnetic field $B_0$ applied to the DAI slab can induce a charge polarization given by $P_{\theta}(t)=\phi\theta(t)e^2/h$~\cite{Li2022identifying}, where $e$ is the electron charge, $h$ is the Planck constant, and $\phi=B_0S$ ($S$ the area of the DAI normal to $B_0$) is the magnetic flux penetrating the DAI. This charge polarization should also evolves adiabatically as the time-dependent axion field $\theta(t)$, leading to an equivalent magnetoelectric current. At particular instant of time, the unbalanced charge distribution illustrated in Fig.~\ref{fig2}a can be calculated by using the lattice Green's functions provided in Methods. Figure~\ref{fig2}b displays the unbalanced charge distribution induced by a static magnetic field along $\hat{z}$-direction at $t_1=0$ for a system with size $L_y=16$, $L_z=10$ and $A_x=0.1$, which corresponds to the charge polarization originating from the static axion term $\theta_0$. The charge distributions corresponding to $\delta\theta(t)$ are further plotted in Fig.~\ref{fig2}c ranging from $t=0$ to $t=\pi/2\omega$ with an interval of $\pi/10\omega$ after subtracting the background charge distribution displayed in Fig.~\ref{fig2}b. 

\noindent  Moreover, the red inverted triangle in Fig.~\ref{fig2}d shows the instant charge polarization given by $\delta P(t)=[\delta Q(z=-1/2)-\delta Q(z=1/2)]/2$, whose temporal evolution is also a harmonic function with a doubled frequency, in agreement with the dynamic axion field shown in Fig.~\ref{fig1}e. The time derivative of this instant charge polarization gives the dynamic magnetoelectric current, which is represented by the blue triangle shown in the same figure. The black dashed line is the anticipated magnetoelectric current obtained from the time derivative of the dynamic axion field at $A_x=1$ shown in Fig.~\ref{fig1}e, and is plotted in the same figure as a benchmark. Despite the slit deviation between the two results, which can be ascribed to the finite size effect, both data can be well fitted by using a double-frequency harmonic function. This quantitative expression of the magnetoelectric current is further verified by the fast-Fourier transform shown in Fig.~\ref{fig2}e.
Since this harmonic current originates directly from the axion dynamics in DAI driven by the AFMR, it can thus stand as an evidence for identifying DAIs. Nevertheless, in view of the high frequency of AFMR, which can even reach terahertz, quantitatively detecting a temporal current with such a high frequency remains a challenge. 

\begin{figure}[t]
  \centering
  \includegraphics[width=1.0\textwidth]{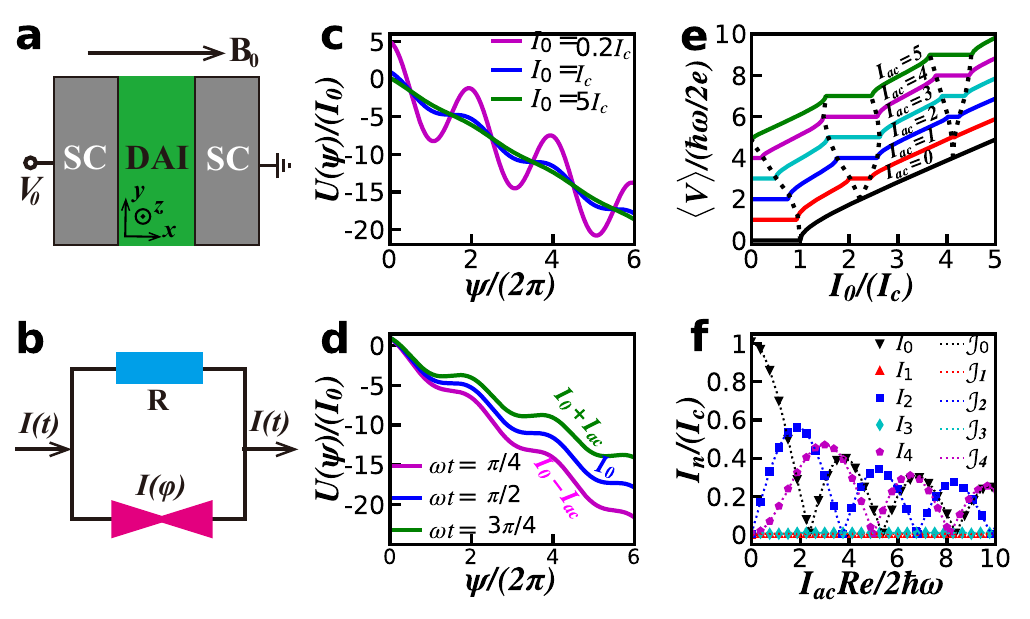}
  \caption{\textbf{Transport properties of a DAI Josephson junction.}
	\textbf{a} Schematic for a Josephson junction with a DAI sandwiched between two identical superconductors. A DC bias $V_0$ (with the right superconductor grounded) along with a static magnetic field $B_0$ are applied across the Josephson junction.
	\textbf{b} The resistively shunted junction model with an electrical resistance $\bf{R}$ shunted with the Josephson junction $\bf{I}(\psi)$. $I(t)$ is the total applied current.
	\textbf{c} and \textbf{d} are static and dynamic washboard potential versus the phase difference $\psi$.
	\textbf{e} Time-averaging voltage output as a function of the electron current input (proportional to the DC bias $V_0$) under different $I_{ac}$ (in unit of $2\hbar\omega/Re$). Here, the datas are vertical shifted one by one for clarity, and the black dashed lines that mark the boundary of each Shapiro steps are the guide to the eye. 
	\textbf{f} Shapiro step widths $I_n$ as a function of $I_{ac}$. The markers here are obtained from the RSJ model while the dashed lines of the same color are fitted by using the first kind Bessel function in Eq.~\ref{super_curr}.
		  	}
\label{fig3}
\end{figure}

\noindent\textbf{Doubled Shapiro steps in the DAI Josephson junction}\\
\noindent To detect this fast dynamic magnetoelectric current unique to the axion dynamics in DAI, we consider a Josephson junction consisting of a DAI sandwiched between two identical superconductors, as schematically shown in Fig.~\ref{fig3}a. Through Andreev reflection, the dynamic magnetoelectric current could stimulate a coherent phase difference across the junction, potentially synchronizing with the AC Josephson current when subjected to additional DC bias and hence becoming a static transport signals termed Shapiro steps. To explore the transport properties of such a DAI Josephson junction, we employ the resistively shunted junction model illustrated in Fig.~\ref{fig3}b, where the system is simulated as an effective circuit featuring a resistance $R$ in parallel to the Josephson junction~\cite{tinkham2004introduction}. This model captures not only the tunneling electron pairs but also the leakage currents through the insulating barrier at the interfaces. The total current then comprises two components: one is the normal supercurrent $I_0$ adjustable by applied DC bias $V_0$ while the other is the time-dependent  axion current $I_\theta(t)$ induced by a static magnetic field $B_0$. Moreover, this current can be further expressed as the sum of the parallel currents in the Kirchhoff's circuit, whose dynamics can be quantitatively described by the equation of motion~\cite{heiselberg2013shapiro,stewart1968current} 
 \begin{align}\label{RSJ}
 I_0+I_{\theta}(t)=\frac{V}{R}+I_c\sin{\psi},
 \end{align}
where $I_0+I_{\theta}(t)$ is applied current, $V=\hbar\dot{\psi}/2e$ is the effective voltage with $\psi$ the phase difference across the Josephson junction and $I_c$ is the critical supercurrent. This equation of motion can be solved numerically by using the Runge-Kutta method after substituting the dynamic current $I_{\theta}(t)=-I_{ac}\sin{2\omega t}$, which is provided in Methods.

\noindent We first rewrite Eq.~\ref{RSJ} as $\hbar\dot{\psi}/2eR=-\partial{U(\psi,t)}/\partial{\psi}$ where $U(\psi,t)=-[I_0+I_\theta(t)]\psi-I_c\cos{\psi}$ is the washboard potential. In the absence of dynamic axion current $I_\theta(t)$, the static washboard potential for different $I_0$ displayed in Fig.~\ref{fig3}c is generally consistent with that in conventional Josephson junctions. When $I_0<I_c$, the system is trapped in a local minimum maintaining a constant phase. Increasing the current to $I_0=I_c$ leads to an instability responsible for the emergence of phase oscillation. The static washboard potential apparently obeys a linear Ohmic relation $I_0=V/R$ when $I_0\gg I_c$. The dynamic washboard potential for non-vanishing $I_\theta(t)$ is presented in Fig.~\ref{fig3}d, which oscillates between two critical boundaries represented by the green and magenta lines. 
This washboard potential exhibits a halved periodicity with $T=\pi/\omega$, in sharp contrast to conventional $T=2\pi/\omega$. During one period, the system remains stable as long as the phase changes $2n\pi$ with $n$ an integer, resulting in a velocity $\dot{\psi}=2n\pi/T=2n\omega$ in the DAI Josephson junction. 
As a result, the Shapiro steps emerge when the motion of the system coincides with the dynamics of the washboard potential , manifested as $\omega_0(=2eV_0)=2n\omega$.
Indeed, this washboard potential is instructive to understand the behavior of the $I$-$\langle V\rangle$ curve.

\noindent Figure~\ref{fig3}e shows the $I$-$\langle V\rangle$ curve with different $I_{ac}$ that is proportional to the magnetic flux $\phi$ and thus $B_0$. When the magnetic field is absent, $I_{ac}=0$ and a typical $I$-$\langle V\rangle$ curve for conventional Josephson junctions without any Shapiro steps is restored as the dynamic axion current is vanishing [$I_\theta(t)=0$]. In the presence of magnetic field ($I_{ac}\ne0$), we see that the even Shapiro steps emerge at $\langle V\rangle=\hbar\omega/e$ while the odd steps disappear completely, in agreement with the halved periodicity in the dynamic washboard potential above. 
Because the voltage drop inside one Shapiro step is zero, the magnitude, also referred to as Shapiro spike, can be obtained through the identification of the width of the plateaus at $\langle V\rangle=\hbar\omega/2e$, which is depicted by colorful markers in Fig.~\ref{fig3}f as a function of $I_{ac}$. The result reaffirms that only doubled Shapiro steps that evolve in perfect accordance to the Bessel functions sustain in the DAI Josephson junction.

\noindent\textbf{Phenomenological scenario}\\
\noindent To understand these doubled Shapiro steps heuristically, we provide a phenomenological analysis following Barone and Patern\`o's argument~\cite{John1982}. 
When the DAI is intimately connected to two identical superconductors, hence forming a DAI Josephson junction, the magnetoelectric current induced by applied magnetic field $B_0$ leads to an effective bias voltage $V_{DAI}(t)=I_{\theta}(t)R$ across the Josephson junction, where $R$ is the effective resistance originating from the leakage current at the interfaces between the DAI and the superconductors. This voltage in turn populates a time-dependent phase difference given by Josephson's second equation $\Phi_{\theta}=\int{dt} V_{DAI}(t)2e/\hbar$, which is $\Phi_{\theta}=eV_1\cos{2\omega t}/\hbar\omega$ with the effective voltage $V_1=I_{ac}R$. The total phase difference across the junction, combined with an additional DC bias $V_0$, can thus be expressed as $\Phi=2eV_0t/\hbar+eV_1\cos{2\omega t}/\hbar\omega$. The Josephson current under such a phase difference thus reads $I_s(t)=I_c\textrm{Im}\{\exp{[i(2eV_0t/\hbar+eV_1\cos{2\omega t}/\hbar\omega)]}\}$ where $I_c$ is the critical supercurrent.  Applying the Jacobi-Anger expansion recasts the Josephson current as~\cite{gradshteyn2014table}
\begin{align}\label{super_curr}
I_s(t)=I_c\sum_{n}\mathcal{J}_n(eV_1/\hbar\omega)\sin{[(2eV_0/\hbar+2n\omega)t]},
\end{align}
where $\mathcal{J}_n(x)$ is the first kind Bessel function.  Equation~\ref{super_curr} manifestly shows that the supercurrent oscillates as a function of time unless at $eV_0=-n\hbar\omega$, where doubled Shapiro steps appear with a magnitude $I_s^n=I_c\mathcal{J}_n(eV_1/\hbar\omega)$. To check the consistency independently, we calculate the Shapiro steps as a function of $V_1$ by using Eq.~\ref{super_curr} and superimpose the result as a dotted line in the same figure. We see that the two results agree with each other remarkably well, which demonstrates the consistency and reliability of obtained results. 

\begin{figure}[t]
  \centering
  \includegraphics[width=0.7\textwidth]{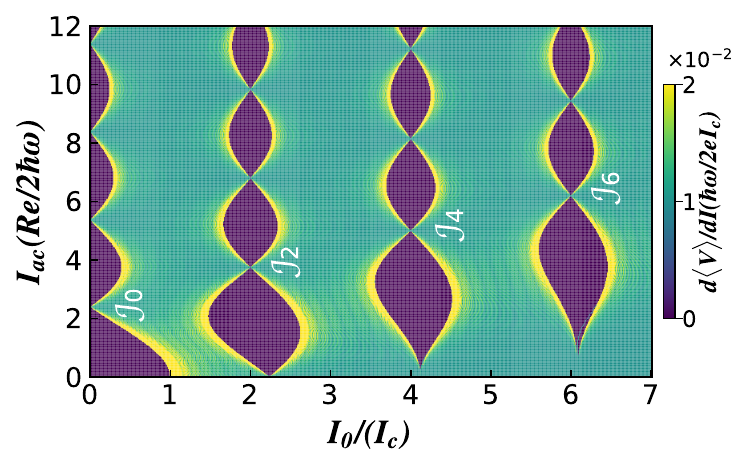}
  \caption{\textbf{Differential conductance of the DAI Josephson junction.}
	Differential conductance $dV/dI$ on the $I_0-I_{ac}$ plane. $I_0$ is proportional to applied DC voltage $V_0$ while $I_{ac}$ is proportional to applied static magnetic field $B_0$. $\mathcal{J}_{\alpha}$ represents the first kind Bessel function of order $\alpha$.
		  	}
\label{fig4}
\end{figure}

\noindent Finally, we determine the differential conductance $d\langle V\rangle/dI$ on the $I_0$-$I_{ac}$ plane by using the resistively shunted junction simulation and show the result in Fig.~\ref{fig4}. We observe a clear and sharp double Shapiro step pattern, where the differential conductance plateaus appear only near the location $I_0=2nI_c$. The boundaries of the plateaus marking the width of the doubled Shapiro steps coincides quantitatively with the Bessel function given in Eq.~\ref{super_curr}.
This unique differential conductance pattern provides a fingerprint of the DAIs and meanwhile, offers a platform to simulate the axion electrodynamics in condensed matter systems.

\noindent\textbf{Discussion}\\
\noindent Despite that a microwave is indispensable in both conventional Josephson junction and the DAI Josephson junction discussed in this work, we stress that the underlying mechanisms of the doubled Shapiro steps are completely different. In conventional Josephson junctions, microwave plays the role of a driven electric field used to induces a harmonic phase difference resonating with an additional DC bias coherently, resulting in a typical Shapiro steps appear consecutively at $\langle V\rangle=\hbar\omega/2e$. In the sharp contrary, the microwave here is solely used to populate AFMR. The doubled Shapiro steps appear at $\langle V\rangle=\hbar\omega/e$ originate from the coherence between the phase difference induced simultaneously from applied DC bias and the dynamic magnetoelectric current, which can thus be further ascribed to the axion dynamics inside DAIs. 

\noindent On the other hand, performing Euler-Lagrangian equation to $\mathcal{L}_{\theta}$ yields the following classical equation of motion for the dynamic axion field~\cite{Sekine2021axion,Beck2013possible}
\begin{align}\label{deom}
\ddot{\delta\theta}-\nabla^2\delta\theta+m_{\theta}^2\delta\theta=\alpha{\bf{E}}\cdot{\bf{B}},
\end{align}
where $m_\theta$ is the axion mass. Although there are ongoing debates regarding the magnitude of axion mass in condensed matter systems~\cite{Sekine2014axionic,Jan2021axion,Ishiwata2021axion,Ishiwata2022topology}, it is straightforward to see that the dynamic axion field induced by AFMR, $\delta{\theta}(t)=A_\theta\cos{2\omega t}$, constitutes an analytical solution to Eq.~\ref{deom} in a uniform system without any electromagnetic field. From this viewpoint, we obtain the axion mass $m_\theta=2\omega$. Consequently, the doubled Shapiro steps proposed here also provide a method to detect the axion mass.

\noindent In Mn${}_2$Bi${}_2$Te${}_5$, as the anisotropy and the exchange interaction are $H_A=0.8$meV and $H_E=0.1$meV~\cite{Li2020intrinsic}, respectively,  the intrinsic AFMR frequency identical to that of the static magnetic field required to drive this AFMR is $f=\omega/2\pi\approx 143$GHz. So the AFMR in Mn${}_2$Bi${}_2$Te${}_5$ can be accomplished through a linearly polarized sub-terahertz radiation analogous to that in Cr${}_2$O${}_3$~\cite{Li2020} and MnF${}_2$~\cite{Priyanka2020sub}. Moreover, the typical tilting angle of the N\'eel vector under AFMR is about $A_x\sim 1\%$, in which case the amplitude of the dynamic axion field is $A_\theta\sim1.0\times 10^{-5}$. For a DAI Josephson junction with size $S=10^{-6}$ m$^2$ and contact resistance $R\sim10\Omega$~\cite{Ohkubo2022}, the magnitude of required static magnetic field to observe a sizable doubled Shapiro steps is about $B_0\sim0.02$Tesla. Those can be readily achieved within current experiments. 

\noindent In summary, we have demonstrated a unique doubled Shapiro steps in a superconductor-DAI-superconductor Josephson junction driven by a DC bias and a static magnetic field instead of a microwave. We ascertain that this distinctive phenomenon arises from the frequency doubling of the axion electrodynamics in DAI under the driven of a linearly polarized microwave, or alternatively from the coherent resonance between the axion mass and applied DC bias, which thus provides a fingerprint of the DAIs and can also be utilized  to evaluate the mass term of axion field in antiferromagnetic topological insulators.

\noindent\textbf{Methods}\\
\noindent \textbf{Lattice Hamiltonian of the DAI.} The low-energy effective Hamiltonian presented in Eq.~\ref{DAI_Ham} can be discretized by using the $k\cdot p$ theory. Performing the substitutions $k_{\alpha=x,y,z}=-i\partial_{\alpha}\rightarrow -i(\psi_{{\bf{i}}+\alpha}-\psi_{{\bf{i}}})/(2a_0)$ and $k_{\alpha}^2=-\partial_{\alpha}^2\rightarrow -(\psi_{{\bf{i}}+\alpha}+\psi_{{\bf{i}}-\alpha}-2\psi_{\bf{i}})/a_0^2$ with $a_0$ the lattice constant, we can write the effective Hamiltonian as
\begin{align}\label{Lattice_Hamiltonian}
\mathcal{H}(t)=\sum_{\bf{i}}\psi_{\bf{i}}^{\dagger}T_{\bf{i}}(t)\psi_{\bf{i}}+(\psi_{\bf{i}}^{\dagger}T_{x}\psi_{{\bf{i}}+x}+\psi_{\bf{i}}^{\dagger}T_{y}\psi_{{\bf{i}}+y}+\psi_{\bf{i}}^{\dagger}T_{z}\psi_{{\bf{i}}+z}+\textrm{H.c.}),
\end{align}
where $\textrm{H.c.}$ is the shorthand for the Hermitian conjugate, and the hoping matrices take the following form
\begin{align}\label{hoping_matrix}
\begin{split}
&T_{\bf{i}}=(m_0+6B/a_0^2)\Gamma_4+m_5n_z(t)\Gamma_5+m_5n_x(t)\Gamma_1\\
&T_x=-iA/(2a_0)\Gamma_1-B/a_0^2\Gamma_4\\
&T_y=-iA/(2a_0)\Gamma_2-B/a_0^2\Gamma_4\\
&T_z=-iA/(2a_0)\Gamma_3-B/a_0^2\Gamma_4,
\end{split}
\end{align}
with $n_x(t)=A_x\sin{\omega t}$ and $n_z(t)=\sqrt{1-A_x^2\sin^2{\omega t}}$. Note that in Eq.\ref{Lattice_Hamiltonian} the translation symmetry along all three spatial directions in the Cartesian coordinate is well preserved. 

\noindent In the presence of a static magnetic field ${\bf{B}}=\begin{pmatrix}0,&0,&B\end{pmatrix}$ along $\hat{z}$-direction, the Hamiltonian in Eq.~\ref{Lattice_Hamiltonian} acquires a Peierls phase $\phi_{\bf{ij}}=2\pi\int_{\bf{i}}^{\bf{j}}{\bf{A}}\cdot d{\bf{r}}/\phi_0$ inside one unit cell from site $\bf{i}$ to $\bf{j}$, where $\bf{A}$ is the vector potential and $\phi_0=h/e$ is the magnetic flux quanta. In particular, we use the Landau gauge, therefore ${\bf{A}}=\begin{pmatrix}-yB,&0,&0\end{pmatrix}$. The Hamiltonian thus becomes
\begin{align}\label{Lattice_Hamiltonian_B}
\mathcal{H}(t)=\sum_{\bf{i}}\psi_{\bf{i}}^{\dagger}T_{\bf{i}}(t)\psi_{\bf{i}}+[\psi_{\bf{i}}^{\dagger}T_{x}f_x(y)\psi_{{\bf{i}}+x}+\psi_{\bf{i}}^{\dagger}T_{y}\psi_{{\bf{i}}+y}+\psi_{\bf{i}}^{\dagger}T_{z}\psi_{{\bf{i}}+z}+\textrm{H.c.}],
\end{align}
where $f_x(y)=e^{-iyBa_0^2/\phi_0}$.

\noindent \textbf{Green's functions method for calculating the charge polarization and the magnetoelectric current.}
Since the AFMR frequency is orders of magnitude smaller compared to the electron response time, we can treat the time-dependent AFMR adiabatically. In this case, the magnetoelectric current induced by a static magnetic field $\bf{B}$ along $\hat{z}$-direction can be obtained from the time derivative of the instant charge polarization. Subsequently, this instant charge distribution parallel to the magnetic field can be expressed as~\cite{Li2022identifying,Li2024High}
\begin{align}\label{static_charge_distribution}
q(z,t)=\frac{e}{2\pi}\sum_y\int_{\infty}^{\epsilon_f}{d\epsilon}\int{dk_x}\textrm{Im}\{\textrm{Tr}[G^r(\epsilon,t)]\},
\end{align}
where $\epsilon_f$ is the Fermi energy and the Green's function $G^r(\epsilon,t)=[\epsilon+i\eta-\mathcal{H}(t)]^{-1}$ with $\eta$ the imaginary line width function. 

\noindent Moreover, equation~\ref{static_charge_distribution} includes only the contribution from negative electron charge. To obtain the unbalanced charge distribution as well as the instant charge polarization, it is necessary to include the positive background charge originating from ions that compensates the electron charge, which has the form $q_{b}=\sum_{z}q(z,t)/L_z$ owing to the charge conservation. As a result, the unbalanced charge distribution at time $t$ can finally be expressed as $Q(z,t)=q(z,t)+q_b$.

\noindent \textbf{Fourth order Runge-Kutta method for solving the equation of motion of the DAI Josephson junction.} For convenience, we rewritten Eq.~\ref{RSJ} in the form $\dot{\psi}=f(t,\psi)$ with $f(t,\psi)=2I_cRe(I_0/I_c-I_{ac}\sin{2\omega t}/I_c-\sin{\psi})/\hbar$. This is a first-order differential equation that can be numerically solved by using the typical Runge-Kutta method. For a step size $h>0$, the phase difference $\psi$ can be obtained iteratively by using the following equations
\begin{align}\label{RK_equation}
\begin{split}
\psi(t_{n+1})&=\psi(t_{n})+\frac{h}{6}(k_1+2k_2+2k_3+k_4),\\
t_{n+1}&=t_n+h,
\end{split}
\end{align}
where
\begin{align}\label{RK_coeffcient}
\begin{split}
k_1&=f(t_n,\psi(t_n)),\\
k_2&=f(t_n+h/2,\psi(t_n)+hk_1/2),\\
k_3&=f(t_n+h/2,\psi(t_n)+hk_2/2),\\
k_4&=f(t_n+h,\psi(t_n)+hk_3).
\end{split}
\end{align}
The equation of motion can then be solved numerically by substituting the initial value $\psi(t=0)=0$ and $\dot{\psi}(t=0)=0$. 
The average voltage across the Josephson junction, as a function of the existing parameters $I_0$ and $I_{ac}$ that are adjustable through the manipulation of $V_0$ and $B_0$, respectively, is then given by Josephson's second equation $V=\int_{t}^{t+T}dt\hbar\dot{\psi}/2eT$ with the period $T=\pi/\omega$.

\noindent\textbf{Data availability}\\
  	The data that support the plots within this paper and other findings of this study are available from the corresponding authors upon request.

    \noindent\textbf{Code availability}\\
  	The code deemed central to the conclusions is available from the corresponding authors upon request..

    \noindent\textbf{References}
    \bibliographystyle{naturemag}
    \bibliography{ref}
    \noindent\textbf{Acknowledgements}\\
    H.J. acknowledges the support from the National Key R\&D Program of China (Grants No. 2019YFA0308403 and No. 2022YFA1403700) and the National Natural Science Foundation of China (Grant No. 12350401). Y.-H.L. acknowledges the support from the Fundamental Research Funds for the Central Universities. X.C.X. acknowledges the support from the Innovation Program for Quantum Science and Technology (Grant No. 2021ZD0302400). R.C. acknowledges the support from the AFOSR (Grant No. FA9550-19-1-0307). Y.-H.L. is also grateful for the financial support from the State Key Laboratory of the Surface Physics and the Department of Physics at Fudan University.
    
    \noindent\textbf{Author contributions}\\
    Y.-H.L, H.J and X.C.X conceived the initial idea of doubled Shapiro steps in the dynamic axion insulator Josephson junction. Y.-H.L performed calculations with assistance from Z.Z. Y.-H.L. wrote the manuscript with contributions from all authors. H.J. and X.C.X. supervised the project.
    
    \noindent\textbf{Competing interests}\\
    The authors declare no competing interests.
    
    \noindent\textbf{Additional information}\\ 
    \textbf{Supplementary information} The online version contains supplementary material available at .\\
    \textbf{Correspondence} and requests for materials should be addressed to H.J. or X.C.X.

    \setcounter{figure}{0}
    \captionsetup[figure]{labelfont={bf}, name={Extended Data Fig.}, labelsep=period}

\end{document}